\pdfoutput = 1

\documentclass[%
 aip,
 jmp,%
 amsmath,amssymb,
 reprint,%
]{revtex4-1}

\usepackage{graphicx}
\usepackage{dcolumn}
\usepackage{bm}
\usepackage{wasysym}
\usepackage{hyperref}
\usepackage{algorithm}
\usepackage{algpseudocode}
\usepackage{ifpdf}
\usepackage{lineno}
\usepackage{amsmath}
\usepackage{amssymb}
\usepackage{booktabs}
\usepackage{dsfont}
\usepackage{bbm}
\usepackage[nearskip,caption=false,font=footnotesize]{subfig}

\usepackage{color}
\usepackage{multirow}
\definecolor{greencolor}{rgb}{0,0.5,0.2}
\definecolor{redcolor}{rgb}{0.,0.,0.}
\definecolor{bluecolor}{rgb}{0.,0.,0.}
\definecolor{greycolor}{rgb}{.5,.5,.5}
\def\Red#1{{\color{redcolor} #1}}

\def\Blue#1{{\color{bluecolor} #1}}

\newtheorem{definition}{Definition}

\newcommand{\indicator}[1]{\mathbbm{1}_{\left[ {#1} \right] }}

\begin{document}

\preprint{AIP/123-QED}

\title[Competitive Learning Approach to Disambiguation in Collaborative Networks]{Network-Based Stochastic Competitive Learning Approach to Disambiguation in Collaborative Networks}

\author{Thiago Christiano Silva}
\email{thiagoch@icmc.usp.br}
\affiliation{
Institute of Mathematics and Computer Science, University of S\~{a}o Paulo \\
S\~{a}o Carlos, S\~{a}o Paulo, P. O. Box 369, 13560-970, Brazil
}


\author{Diego Raphael Amancio}
\email{diego.amancio@usp.br}
\affiliation{
Institute of Physics of S\~{a}o Carlos, University of S\~{a}o Paulo \\
S\~{a}o Carlos, S\~{a}o Paulo, P. O. Box 369, 13560-970, Brazil
}%

\date{\today}

\begin{abstract}
Many patterns have been uncovered in complex systems through the application of concepts and methodologies of complex networks. Unfortunately, the validity and accuracy of the unveiled patterns are strongly dependent on the amount of unavoidable noise pervading the data, such as the presence of homonymous individuals in social networks. In the current paper, we investigate the problem of name disambiguation in collaborative networks, \Red{a task that plays a fundamental role on a myriad of scientific contexts. In special, we use an unsupervised technique which relies on a particle competition mechanism in a networked environment to detect the clusters. It has been shown that, in this kind of environment, the learning process can be improved because the network representation of data can capture topological features of the input data set.
Specifically, in the proposed disambiguating model, a set of particles is randomly spawned into the nodes constituting the network.} As time progresses, the particles employ a movement strategy composed of a probabilistic convex mixture of random and preferential walking policies. \Red{In the former, the walking rule exclusively depends on the topology of the network and is responsible for the exploratory behavior of the particles. In the latter, the walking rule depends both on the topology and the domination levels that the particles impose on the neighboring nodes. This type of behavior compels the particles to perform a defensive strategy, because it will force them to revisit nodes that are already dominated by them, rather than exploring rival territories.
Computer simulations conducted on the networks extracted from the arXiv repository of preprint papers and also from other databases reveal the effectiveness of the model, which turned out to be more accurate than traditional clustering methods.}
\end{abstract}

\maketitle

%

\noindent
{\bf Complex networks concepts have been employed in a myriad of contexts to model real systems. In the current paper, we use the complex network framework to address the problem of disambiguating authors' names in scientific manuscripts. While traditional strategies are based only on the recurrence of collaborators, we approach the task with a stochastic model based on the connectivity patterns in the collaborative network. The discriminability observed in three distinct data sets of preprint papers revealed the effectiveness of the model, which is significantly more precise than other competing systems.
}

\section{Introduction}

For any piece of work available in the literature, a fundamental issue concerns the identification of the respective author(s). Among several reasons, the recognition of authorship in manuscripts plays a prominent role in the scientific context, where researchers might be interested in identifying potential collaborators. Despite its apparent simplicity, authorship identification still represents an unsolved task for information sciences~\cite{information}. Difficulties arise, for example, when authors' names display variant forms, when spelling errors are made or even when names change due to marriage. One of the most common problems occurs when multiple authors share the very same name or alias, hampering the credibility of applications dependent on the accurate authorship identification.  For example, the hasty choice of researchers for refereeing papers or the inaccurate quantification of researchers' merit based on their publication profile might undermine the efficiency of the system as a whole. In order to minimize the problems stemming from  the presence of ambiguities in authors' names, many scholars and publishers have called for more efficient disambiguation algorithms~\cite{information}.

	Traditional methods for discriminating ambiguous names in the scientific context are based on the patterns of collaboration \cite{Thiago2012f}, on the analysis of metadata~\cite{song2007} and on the content of papers~\cite{Hofmann}. One of the simplest approaches consists in analyzing collaborative networks, where authors appear linked when they collaborate together in at least one paper. The success of this approach can be explained by the emergence of collaboration patterns characterizing homonymous authors~\cite{newmanpnas}. \Red{A simple approach assumes the identification of direct collaborators because, in many circumstances, authors whose names are identical pertain to distinct scientific communities.} In the current paper, we address the problem of disambiguating authors' names in the arXiv repository of preprint papers. In special, we devise and adapt an unsupervised strategy where each ambiguous author is characterized by the recurrence of collaborative patterns.
\Red{The proposed algorithm is based on the dynamics of particles performing a walk conditioned by the so-called random and preferential rules.} In particular, we have found a significant improvement of the discrimination efficiency when we compare our technique with traditional pattern recognition methods. \Red{We believe that our results might be useful to the development of better disambiguating systems. In special, we show that the proposed methodology can be straightforwardly applied in more complex types of attributes, such as metadata or textual contents.} Because the devised strategy is generic, it can also be extended to other related problems, which are pervading, \Red{for instance,} in the natural language processing research.

	\textbf{\Blue{This paper is organized as follows. In Section \ref{sec:Network-Formation-Description}, we deal with the methodology used to disambiguate authors in collaborative networks. In Section \ref{sec:Model-Description} we introduce the model based on the competition of particles. In Sections \ref{resultados} and \ref{resultado2}, \Red{we display the results and discussions obtained by the proposed technique.} Finally, in Section \ref{conclusao}, we draw some final conclusions about our work.}
}

\section{Methodology}
\label{sec:Network-Formation-Description}

In this section, we detail how the relationships are taken into account in the process of building the network, i.e., the collaborative network. As a swift remark, our objective is to encounter the different entities in the network, here represented by the same authors. Note that an entity may be represented by one or more nodes. The task is to disambiguate such matter and, therefore, discover whose nodes represent the same author. For that, we are given a similarity matrix of all the nodes in the network.
In such matrix, it is included intermediate nodes (authors) that we do not desire to disambiguate, whose sole purpose is to help in grouping the desired nodes. \Red{With the aid of these intermediate nodes in the network, we are going to employ a measure based on the \emph{passage time}, which is a concept borrowed from the Markov Chain Theory \cite{Cinlar1975}, to calibrate and set all the edge weights in the network}. Having established the edge weights, we perform a network reduction in the following manner: we deliberately remove all the intermediate nodes from the network. In the final reduced network, we apply our competitive learning algorithm pertaining to the unsupervised scheme in order to find the clusters in the reduced network. \Red{We hope that each cluster will contain all the representative nodes of same entity, i.e., the same author. In the next subsections, we describe how the collaborative network is built in a detailed manner.}

\subsection{Collaborative Network Formation}

\begin{figure}
		\begin{center}
		\includegraphics[width=0.35\textwidth]{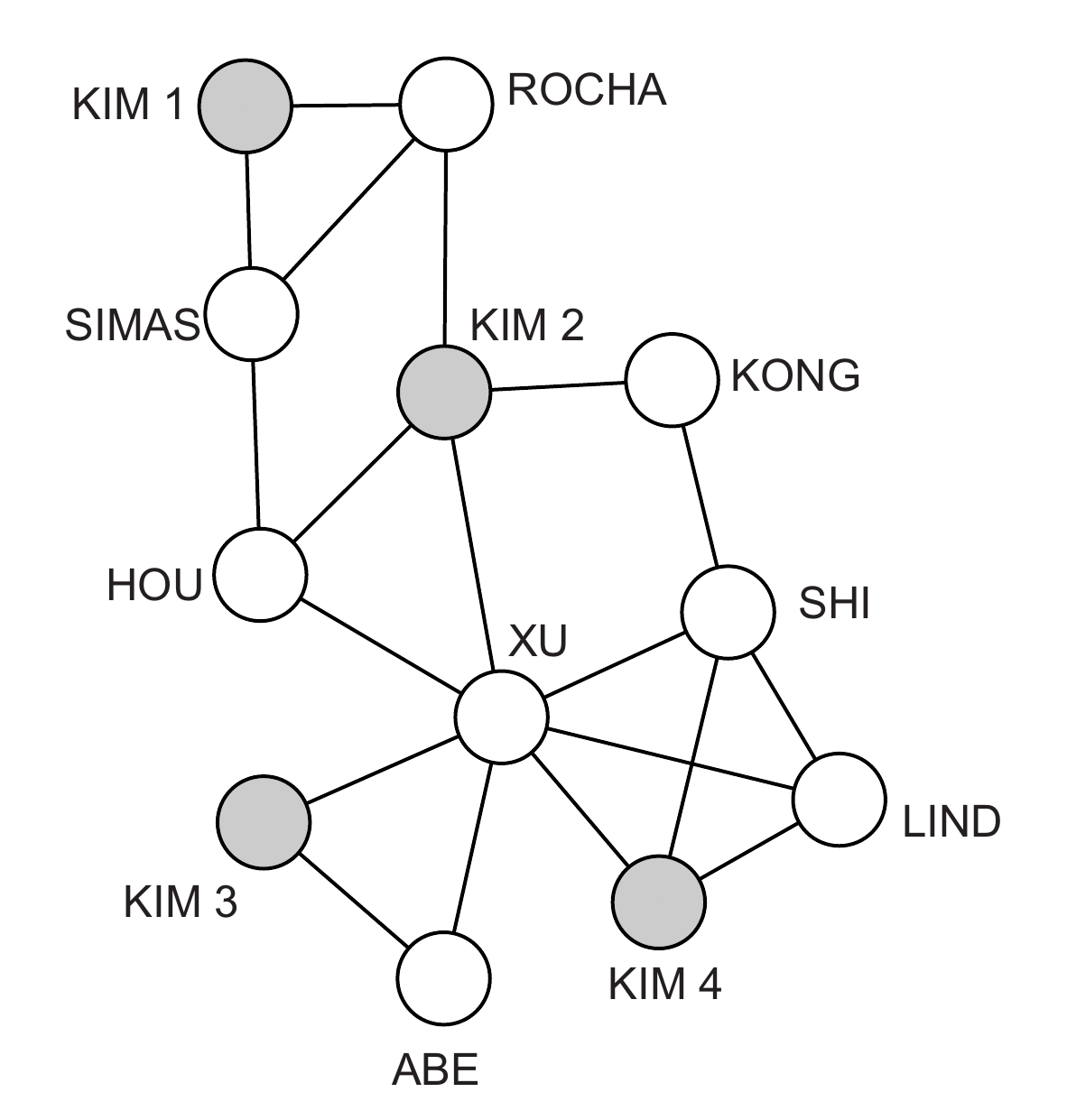}
		\end{center}
		\caption[\it]{\it Example of a collaborative network for a toy database. The papers considered are: paper 1 (Kim, Rocha and Simas), paper 2 (Kim, Xu and Abe), paper 3 (Kim, Xu and Lind), paper 4 (Kim, Hou and Xu), paper 5 (Kim and Rocha), paper 6 (Kim and Kong), paper 7 (Simas and Hou), paper 8 (Kong and Shi), paper 9 (Shi and Kong), and paper 10 (Lind, Xu and Shi). While white nodes represent auxiliary nodes, gray nodes depict those ones whose associated label is a ambiguous name. If two names appear in the same paper, they are linked with each other in the network.}
		\label{fig1}
\end{figure}

To capture the relationship between authors, a collaborative network is created. In particular, each distinct author's name is represented in the collaborative network as a node. Edges are established between two nodes if they co-occur in at least one of the articles. To illustrate the construction of the network, consider the database listed in the caption of Fig. \ref{fig1}. Note that two authors who have published collectively at least one article (see e.g. ``Shi''  and ``Kong'' in paper 8) are connected in the respective network. In particular, for this toy database, we intend to disambiguate the various observations of the name ``Kim.'' Since it is desirable to associate the observations to the same entity, each of the ambiguous name observations of ``Kim'' generates a distinct node (``Kim 1'', ``Kim 2'', ``Kim 3'' and ``Kim 4''). The strength of the links between two nodes $i$ and $j$ is given by the weight:
\begin{equation}
	w_{ij} = \sum_{k \in P} \frac{ \delta_{ijk} }{ |k| },
    \label{eq:formulaDiego}
\end{equation}
where $P$ represents the set of all papers in the database and $\delta_{ijk}$ = 1 provided that authors $i$ and $j$ appear in the same paper $k$ and $\delta_{ijk}$ = 0, otherwise. Note that a divisive factor $|k|$ is included in the expression. This extenuatory term represents the number of authors in paper $k$ and is used to model the effect that relationships involving few authors are usually stronger than those encompassing several authors. Even though weights are not illustrated in Fig. \ref{fig1}, their calculation are straightforward. For example, the weight connecting ``Rocha'' and ``Simas'' is 1/3, while the weight linking ``Kong`` and ``Shi'' is 1/2 (from paper 8) plus 1/2 (from paper 9). The same reasoning can be applied for the remaining edges.

With the aid of Eq. (\ref{eq:formulaDiego}), one can build a network as represented in Fig. \ref{fig1}. \Red{However, if we were to take into account such approach to find the nodes that represent the same author through a clustering task, such measure would not translate, in a reliable manner, the connections among the authors that have co-authored with only distinct persons in each of his/her papers, i.e., nodes that represent the same entity could be situated far away from each other, which is undesirable in a clustering task. By virtue of that, we can say that the construction of the collaborative network with the assistance of Eq. (\ref{eq:formulaDiego}) may only capture local features of the network.} Hence, it would be unable to hold the semantic characteristics of the data in a global fashion manner. With that in mind, we propose a truncated version of the well-known measure \emph{passage time}, which pertains to the Markov Chain Theory. Before going any further, it is worth giving a brief overview of how to derive the proposed measure for the construction of the network.

\subsection{Description of the Similarity Measure}

In this section, we present the proposed similarity measure which will be used when we deal with the application of author's name disambiguation. \Red{First, we show the classical concepts of the Markov Chain Theory.} Then, the proposed measure \emph{per se} is introduced.

\subsubsection{Classical Concepts of the Markov Chain Theory}

In order to use the network-based community detection technique, which will be explained in Section \ref{sec:Model-Description}, we are required to construct a network that represents the data relationships in a satisfactory manner. \Red{In an attempt to do so, we will make use of a well-known measure of the Markov Chain Theory entitled \emph{passage time}. We now give a formal definition of a discrete markov chain in details.}

Let $\Omega$ be a sample space and $P$ a probability measure associated to it. Consider a stochastic process $X = \{X_{t}; t \in \mathbb{N}\}$ with a countable state space $V$; i.e., for each $t \in N = \{0,1,\hdots\}$ and $\omega \in \Omega$, $X_{t}(\omega) \in V$. In other words, $X_{n}$ represents in which node of the network the stochastic process $X$ is at time $t$. In the following, we formalize these concepts.

\begin{definition}
    \emph{\textbf{(Discrete time Markov Chain (MC))}} \cite{Cinlar1975}. The stochastic process $X = \{X_{t}; t \in \mathbb{N}\}$ is called a Markov chain of first order provided that:
    \begin{align}
        P\{X_{t+1} = j \mid X_{0},\hdots,X_{X_{t}}\} = P\{X_{t+1} = j \mid X_{t}\},
    \end{align}
    \noindent $\forall j \in V$ and $t \in \mathbb{N}$. 
\end{definition}

A random walk on a $\mathrm{MC}$ can be defined as follows: a random walker starts
in a state $v$ according to the initial distribution $p_{0}$. \Red{Next, it moves to some state
$v' \in V$ according to the transition probability matrix $P$, which is given. At each time step,
the walker visits a specific node $v \in V$ in the network.} The passage time function precisely counts the number of times
a given node has been visited during a random walk. Next, this notion is elucidated.

\begin{definition}
    \emph{\textbf{(Passage Time)}} \cite{Cinlar1975}. The passage time is a function $pt: V \times V \rightarrow \mathbb{N}$ which counts the number of times the Markov chain process has visited a specific node $v \in V$. Mathematically:

    \begin{align}
        pt(v) = |{t \in N \mid X_{t} = v}| = \sum_{t=0}^{\infty}{\mathds{1}_{\{X_{t}(\omega) = v\}}}
    \end{align}

    \noindent $\forall \omega \in \Omega$, where $\mathds{1}_{.}$ yields $1$ if the argument is $\mathrm{true}$ and $0$, otherwise. 
\end{definition}

Note that, by the monotone convergence theorem, each $(i,j)$th-entry of the domain of $pt$ ($V \times V$) is exactly the $(i,j)$th-entry of the potential matrix (or fundamental matrix) $R$ \cite{Cinlar1975}.

\subsubsection{Description of the Proposed Similarity Measure}

As we have stated, the measure given in Eq. ($\ref{eq:formulaDiego}$) can be readily calculated from a set of papers. Some shortfalls of such measure are that: ($\mathrm{i}$) it can only provide similarity between authors in a local manner; ($\mathrm{ii}$) consequently, it may not correctly capture the similarity from authors that may have co-authored with distinct authors in his/her papers. By virtue of that, we propose a measure that can assimilate such matter with the aid of a user-controllable parameter $l$, which accounts for the length of the random walk to be performed. In general terms, for each node in the network, we perform a random walk of length $l$ according to the probability matrix constructed by Eq. ($\ref{eq:formulaDiego}$), counting the number of times all nodes have received a visit by the random walker. We repeat this process $r$ times for each node. \Red{The aggregate number of visits performed by each particle starting at node $v_{s} \in V$ and ending at another node $v_{e} \in V$  will be the edge weight $A(v_{s},v_{e})$ of the resulting network.} That is, nodes whose distances are greater than the threshold $l$ will automatically have their edge weights set to zero. The parameter $l$ calibration can be conceived as a mechanism of capturing from local to global characteristics of the original network. \Red{As $l$ grows, more global features are taken into account.} An efficient way of computing this measure can be achieved by using stochastic forward variables, similar to those introduced by the Baum-Welch algorithm for Hidden Markov Models~\cite{Baum1970}. Given a state $v \in V$ and a time $t \in \mathbb{N}$, the forward variable $\alpha(v,t)$ determines the probability to reach state $v$ after $t$ time steps. The forward variables, related to a starting node $v_{s}$, are calculated using the following recurrence:

\begin{align}
    \begin{split}
        \mbox{(Case $t = 1$)}\ \ \ & \alpha^{v_{s}}(v,t) = P(v_{s}, v)\\
        \mbox{(Case $t > 1$)}\ \ \ & \alpha^{v_{s}}(v,t) = \sum_{v' \in V}{\alpha(v', t-1)P(v',v)}
    \end{split}
    \label{eq:recurrence}
\end{align}

\noindent where $P(i,j) = w_{ij}$ as indicated in Eq. ($\ref{eq:formulaDiego}$). With the mechanism inherently supplied by Eq. ($\ref{eq:recurrence}$), we expect to find nodes that represent the same entity, but are not directly connected with each other through Eq. ($\ref{eq:formulaDiego}$), \Red{since we will not only cover the direct neighbors of a authors, but all the neighbors within a pre-defined vicinity, which is numerically fixed by the parameter $l$.}

One can see that, for $l = 1$, the method reduces to the special case provided in Eq. ($\ref{eq:formulaDiego}$). For $l > 1$, not only local features (direct neighbors) are taken in consideration, but also neighbor of neighbors and so on. As  $l$ increases, more global features are taken into account in the similarity calculation process. One can see that a critical value of $l_{\mathrm{c}}$ which maximizes the clustering process must exist, because for $l > l_{\mathrm{c}}$, \Red{the global features mix with the local features in a way that the final result becomes compromised in terms of edge weight quality.} A detailed analysis of this $l_{\mathrm{c}}$ is left as future work. Furthermore, for an irreducible (ergodic) and aperiodic Markov chain (network), if $l \rightarrow \infty$, then all the edge weights of the graph approximate to the invariant distribution $\pi$ \cite{Cinlar1975}, i.e., every row of the similarity matrix of the network is equal.

\subsubsection{Network Reduction Method}

Using the similarity measure described in the previous section, we computed all pairs of similarities between entities whose names are ambiguous. Thus, a similarity matrix is obtained to be used as input of the algorithm. Note that at this stage, the nodes of the network that do not represent ambiguous entities are only used to calculate the similarity between entities ambiguous. Thus, they are not part of the aforementioned similarity matrix.


\section{Model Description}
\label{sec:Model-Description}

In this section, the unsupervised particle competition learning model~\cite{Thiago2012a, Thiago2012b, Thiago2012e} is presented.

\subsection{A Brief Overview of the Model}
\label{sec:Brief-Overview}

Consider a graph $\mathcal{G} = \langle \mathcal{V},\mathcal{E} \rangle$, where $\mathcal{V} = \{v_{1},\hdots,v_{V}\}$ is the set of nodes and $\mathcal{E} = \{e_{1},\hdots,e_{L}\} \subset \mathcal{V} \times \mathcal{V}$ is the set of links (or edges). \Red{In the original competitive learning model, a set of particles $\mathcal{K} = \{1,\hdots,K\}$ is inserted into the nodes of the network. The particles are defined as active agents which are able to traverse the network by visiting the vertices of it in agreement with a specific movement policy. Their main purpose is to conquer new vertices by constantly visiting them, while also preventing rival particles from entering and conquering the already dominated vertices. When a particle visits an arbitrary node, it strengthens its own domination level on this node and, simultaneously, weakens the domination levels of all other rival particles on this same node. It is expected that this model, in a broad horizon of time, will end up uncovering the clusters or community in the network in such a way that each particle dominates a cluster or community. It is worth remembering that a community can be conceptualized as a networked representation of a densely subset of vertices interconnected, while a cluster holds the same definition but in a vector-based space (attribute space).}

A particle in this model can be in two states: \textit{active} or \textit{exhausted}. Whenever the particle is active, it
navigates in the network according to a combined behavior of random and preferential walking. The random walking term is responsible
for the adventuring behavior of the particle, i.e., it randomly visits
nodes without taking into account their domination levels. The preferential walking term is responsible for the defensive
behavior of the particle, i.e., it prefers to reinforce its owned territory rather than visiting a node that is not being
dominated by that particle.

So as to make this process suitable, each particle carries an energy term with it. This energy
increases when the particle is visiting an already dominated node by itself, and decreases whenever it visits a
node that is being owned by a rival particle. If this energy drops under a minimum allowed value, the particle becomes exhausted
and is teleported back to a safe ground, \Red{which is the subset of vertices that it is currently dominating. As the authors in Ref. \cite{Thiago2012a} draw attention to, the main idea of introducing this mechanism is to make the model independent of the particles' initial locations. This is rather intuitive in the sense that, given that any particle in the model has a nonzero probability of traversing a sufficient large chain of non dominated vertices, it will eventually get exhausted. Therefore, at the initial stage of the algorithm, where the initial locations of the particle are important, the first trajectories that the particles perform are expected to be reset once they enter the exhausted state. Upon transiting to this state, it will be compelled to get back to its dominated domain no matter how the topology of the network is. Furthermore, the dominated region of each particle is expected to get more stable as time progresses.  When the particle goes back to its domain via the reanimation process,
the exhausted particle will be possibly recharged by visiting the nodes dominated by itself.} In this way, this natural mechanism is responsible for restraining the acting region of each particle and, thus, reduce long-range and redundant visits in the network.

\subsection{The Competitive Transition Matrix}
\label{sec:Competitive-Transition-Matrix}

In this model, each particle $k \in \mathcal{K}$ can perform \Red{two distinct types of movements when it is in the active state:}

\begin{itemize}
    \item \Red{A \emph{random movement term}, modeled by the matrix $\mathbb{P}^{(k)}_{\mathrm{rand}}$;}
    \item \Red{ \emph{preferential movement term}, modeled by the matrix $\mathbb{P}^{(k)}_{\mathrm{pref}}$.}
\end{itemize}

\Red{As we have seen, the two types of movements are orthogonal with regard to their influence on the particles' movement policy. While the random term endows the particles with their defensive behavior, the preferential term gifts them with the exploratory and adventurous features.}

\Red{In order to model such dynamics, consider that  $p(t)=[p^{(1)}(t),p^{(2)}(t),\hdots,p^{(K)}(t)]$ is a stochastic vector, which registers the localization of the set of $K$ particles in the network. In particular, the $k$th-entry, $p^{(k)}(t)$, displays the physical location of particle $k$ at instant $t$. The first strategy in order to build up the competitive system, as the authors in \cite{Thiago2012a, Thiago2012b} indicate, is to find a transition matrix of these particles, i.e., $p(t+1)=[p^{(1)}(t+1), p^{(2)}(t+1),\hdots,p^{(K)}(t+1)]$.}

\Red{Additionally, suppose that $S(t) = [S^{(1)}(t), \hdots, S^{(K)}(t)]$ is a stochastic vector, which keeps track of the current states of all particles at instant $t$. In special, the $k$th-entry, $S^{(k)}(t) \in \{0,1\}$, marks whether the particle $k$ is
active ($S^{(k)}(t) = 0$) or exhausted ($S^{(k)}(t) = 1$) at time $t$. When it is active, the movement policy consists of a combined behavior of randomness and preferential movements. At the hour which it is exhausted, the particle switches its movement policy to a new transition matrix, here referred to as $\mathbb{P}^{(k)}_{\mathrm{rean}}(t)$.} This matrix is responsible for taking the particle back to its dominated territory, in order to reanimate the corresponding particle by recharging its energy (\emph{reanimation
procedure}).

Under these definitions, the transition matrix associated to particle $k$ is defined as:

\begin{align}
    \mathbb{P}^{(k)}_{\mathrm{transition}}(t) &\triangleq (1 - S^{(k)}(t))\left[\lambda \mathbb{P}^{(k)}_{\mathrm{pref}}(t)+(1-\lambda)\mathbb{P}^{(k)}_{\mathrm{rand}}\right]\nonumber\\
    & \qquad + S^{(k)}(t)\mathbb{P}^{(k)}_{\mathrm{rean}}(t),
    \label{eq:transition-matrix-modified}
\end{align}

\noindent where $\lambda \in [0,1]$ counterbalances the fractions of random and preferential movements of particle $k$. In the next, we define the matrices that appear in (\ref{eq:transition-matrix-modified}).

\Red{Firstly, the random matrix only depends on the topology of the network. In this way, this matrix can be fully described once we know the adjacency matrix of the
graph}, which is previously known. In this way, each entry $(i,j) \in \mathcal{V} \times \mathcal{V}$ of the matrix
$\mathbb{P}^{(k)}_{\mathrm{rand}}$ is given by:

\begin{align}
    \label{eq:random-matrix}
    \mathbb{P}^{(k)}_{\mathrm{rand}}(i,j) \triangleq \frac{a_{i,j}}{\sum_{u=1}^{V}{a_{i,u}}},
\end{align}

\noindent where $a_{i,j}$ denotes the $(i,j)$th-entry of the adjacency matrix $A$ of the graph. In short, the probability of an adjacent neighbor $j$ to be visited from  node $i$ is
proportional to the edge weight linking these two nodes.

\Red{Secondly, the preferential matrix depends both on the topology and the domination levels of the particles. The latter is a measure which is calculated using the dynamics of the competitive process itself.} For its definition, it is useful to first define the stochastic vector:

\begin{align}
    \label{eq:number-visits}
    N_{i}(t) \triangleq [N_{i}^{(1)}(t),N_{i}^{(2)}(t),\hdots,N_{i}^{(K)}(t)]^T,
\end{align}

\noindent where $\mathrm{dim}(N_{i}(t)) = K \times 1$, $T$ denotes the transpose operator, and $N_{i}(t)$ stands for the number of visits received by node $i$ up
to time $t$ by all particles scattered throughout the network. Specifically, the $k$th-entry, $N_{i}^{(k)}(t)$, indicates the
number of visits made by particle $k$ to node $i$ up to time $t$.

\Red{Now, we are able to formally define the domination level stochastic vector as:}

\begin{align}
    \label{eq:normalized-number-visits}
    \bar{N}_{i}(t) \triangleq [\bar{N}_{i}^{(1)}(t),\bar{N}_{i}^{(2)}(t),\hdots,\bar{N}_{i}^{(K)}(t)]^T,
\end{align}

\noindent where $\mathrm{\mathrm{dim}}(\bar{N}_{i}(t)) = K \times 1$ and $\bar{N}_{i}(t)$ denotes the relative frequency of
visits of all particles in the network to node $i$ at time $t$. \Red{In particular, the $k$th-entry, $\bar{N}_{i}^{(k)}(t)$,
indicates the relative frequency of visits performed by particle $k$ to node $i$ at time $t$.} Therefore, one has:

\begin{align}
    \label{eq:def-normalized-number-visits}
    \bar{N}_{i}^{(k)}(t) \triangleq \frac{N_{i}^{(k)}(t)}{\sum_{u=1}^{K}{N_{i}^{(u)}(t)}}.
\end{align}

\Red{In view of this, we can define $\mathbb{P}_{\mathrm{pref}}^{(k)}(i,j,t)$, which is the probability of a single particle $k$ to
perform a transition from node $i$ to $j$ at time $t$}, using solely the preferential movement term, as follows:

\begin{align}
    \label{eq:determinist-matrix}
    \mathbb{P}_{\mathrm{pref}}^{(k)}(i,j,t) \triangleq \frac{a_{i,j}\bar{N}_{j}^{(k)}(t)}{\sum_{u=1}^{V}{a_{i,u}\bar{N}_{u}^{(k)}(t)}}.
\end{align}

From (\ref{eq:determinist-matrix}), it can be observed that each particle has a different transition matrix associated
to its preferential movement and that, unlike the matrix related to the random movement, it is time-variant with
dependence on the domination levels of all the nodes ($\bar{N}(t)$) in the network at time $t$. It is worth mentioning that
the approach taken here to characterize the preferential movement of the particles is defined as the visiting frequency of each particle
to a specific node. This means that, as more visits are performed by a particle to a determined node, there will be a higher chance for the same particle to repeatedly visit the same node \cite{Thiago2012a}. Furthermore, it is important to emphasize that
(\ref{eq:determinist-matrix}) produces two distinct features presented by a natural competitive model: (i) the strengthening of
the domination level of the visiting particle on a node; and (ii) the consequent weakening of the domination levels of all other particles on the same node.

\Red{Finally, we define each entry of $\mathbb{P}^{(k)}_{\mathrm{rean}}(t)$ that is responsible for teleporting an exhausted particle $k
\in \mathcal{K}$ back to its dominated territory, in a random manner. This process is performed with the purpose of recharging the particle's energy (reanimation process).} Suppose that particle $k$ is visiting node $i$ when its energy is completely depleted. \Red{In this situation, the particle must regress to an arbitrary node $j$ of its possession at time $t$, according to the following expression:}

\begin{align}
    \mathbb{P}^{(k)}_{\mathrm{rean}}(i, j, t) \triangleq  \frac{\mathds{1}_{\left\{\mbox{arg }\underset{m\in \mathcal{K}}{\operatorname{\mbox{max}}}\left(\bar{N}^{(m)}_{j}(t)\right) = k\right\}}}{\sum_{u = 1}^{V}{\mathds{1}_{\left\{\mbox{arg }\underset{m\in \mathcal{K}}{\operatorname{\mbox{max}}}\left(\bar{N}^{(m)}_{u}(t)\right) = k\right\}}}},
    \label{eq:resurrection-matrix}
\end{align}

\begin{figure}
		\begin{center}
		 \includegraphics[width=0.55\textwidth]{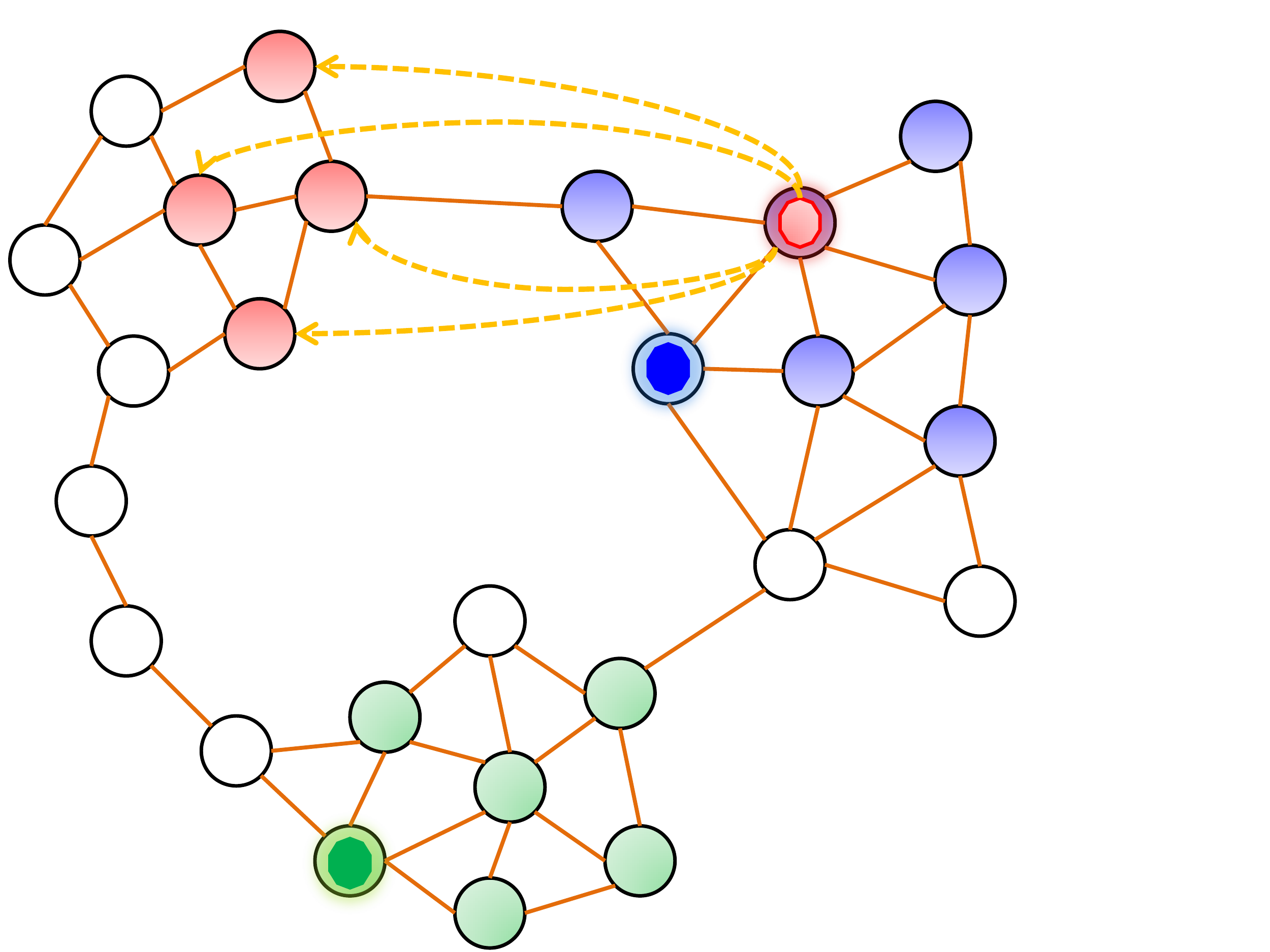}
		\end{center}
		\caption[\it]{\it A reanimation schematic of an exhausted particle. There are three particles in this example: red, blue, and green particles. We fill in the node with the color of the particle which is imposing the highest domination level. Blank nodes represent non dominated nodes. The continuous edges represent the topology of the network, and the dotted lines display the available paths for the exhausted red particle. Since it has become exhausted, note that it will be teleported back to any of its dominated nodes (uniform distribution), regardless of the network topology.}
		\label{fig2}
\end{figure}

\noindent \Red{where $\mathds{1}_{\mathrm{A}}$ is the Heaviside function, which returns $1$ if the logical expression $\mathrm{A}$ is true, and returns $0$, otherwise.} For didactic purposes, Fig. \ref{fig2} portrays a simple scenario of the reanimation procedure taking place. \Red{In this case, the red particle, since it is visiting a node dominated by a rival particle, will have its energy penalized.} Here, we suppose that its energy has been completely depleted and, therefore, the red particle becomes exhausted. Under these circumstances, the reanimation of such particle will occur, which will force the particle to travel back to its dominated territory to be properly recharged.

\Red{With the particles' movement policy fully described, we now discuss the particles' energy update policy. For this end, suppose that $E(t) = [E^{(1)}(t), \hdots, E^{(K)}(t)]$ is a stochastic vector, where the $k$th-entry, $E^{(k)}(t) \in [\omega_{\mathrm{min}},
\omega_{\mathrm{max}}]$, $\omega_{\mathrm{max}} \geq \omega_{\mathrm{min}}$, denotes the energy level of particle $k$ at time
$t$. The limits $\omega_{\mathrm{min}}$ and $\omega_{\mathrm{max}}$ are scalars.} In this scenario, the energy update rule is given by:

\begin{align}
    E^{(k)}(t) =
    \begin{cases}
    \mathrm{min}(\omega_{\mathrm{max}}, E^{(k)}(t-1) + \Delta),& \mbox{if }\mathrm{owner}(k,t) \\
    \mathrm{max}(\omega_{\mathrm{min}},E^{(k)}(t-1) - \Delta),&  \mbox{if }\invneg \mathrm{owner}(k,t)
    \end{cases}
    \label{eq:energy-levels-def}
\end{align}

\noindent where $\mathrm{owner(k,t)} = \left(\mbox{arg }\underset{m \in
\mathcal{K}}{\operatorname{\mbox{max}}}\left(\bar{N}^{(m)}_{p^{(k)}(t)}(t)\right) = k\right)$ is a logical expression that
essentially yields $\mathrm{true}$ if the node that particle $k$ visits at time $t$ (i.e., node $p^{(k)}(t)$) is being
dominated by it, but yields $\mathrm{false}$ otherwise; $\mathrm{dim}(E(t)) = 1 \times K$; $\Delta > 0$ symbolizes
the increment or decrement of energy that each particle receives at time $t$. The first expression in (\ref{eq:energy-levels-def}) represents the increment of the particle's energy and it occurs when particle $k$ visits a node $p^{(k)}(t)$ which is dominated by itself, i.e., $\mbox{arg }\underset{m \in
\mathbb{K}}{\operatorname{\mbox{max}}}\left(\bar{N}^{(m)}_{p^{(k)}(t)}(t)\right) = k$. Similarly, the second expression in
(\ref{eq:energy-levels-def}) indicates the decrement of the particle's energy that happens when it visits a node dominated by rival particles. Therefore, in this model, particles will be given a penalty if they are wandering in rival territory, so
as to minimize aimless navigation of the particles in the network.

Now we advance to the update rule that governs $S(t)$, which is responsible for determining the movement policy of each
particle. As we have stated, an arbitrary particle $k$ will be transported back to its domain only if its energy drops under the
threshold $\omega_{\mathrm{min}}$. Mathematically,  the $k$th-entry of $S(t)$ can be written as:

\begin{align}
    S^{(k)}(t) = \indicator{E^{(k)}(t) \ = \ \omega_{\mathrm{min}}},
    \label{eq:switch-def}
\end{align}

\noindent where $\mathrm{dim}(S(t)) = 1 \times K$. Specifically, $S^{(k)}(t) = 1$ if $E^{(k)}(t) = \omega_{\mathrm{min}}$ and
$0$, otherwise.
The upper limit, $\omega_{\mathrm{max}}$, has been introduced to prevent any
particle in the network from increasing its energy to an undesirably high value and, therefore, taking a long time to become exhausted even if it constantly visits nodes from rival particles. In this way, the community and cluster detection
rates of the proposed technique would be considerably reduced.

\subsection{The Unsupervised Competitive Learning Model}
\label{sec:The-Competitive-Learning-Model}

In light of the results obtained in the previous section, we are ready to enunciate the proposed dynamical system, which models
the competition of particles in a given network. The internal state of the dynamical system is denoted as:

\begin{align}
  \label{eq:state-mod}
  X(t) =
  \left[
    \begin{array}{c}
        p(t)\\
        N(t)\\
        E(t)\\
        S(t)
    \end{array}
  \right],
\end{align}

\noindent and the proposed competitive dynamical system is given by:

\begin{align}
  \label{eq:dynamical-system-mod}
  \begin{split}
      \phi:\left\{
      \begin{array}{l l}
        p^{(k)}(t+1) &= j, \quad j \sim \mathbb{P}^{(k)}_{\mathrm{transition}}(t)\\
        N_{i}^{(k)}(t+1) &=  N_{i}^{(k)}(t) + \indicator{p^{(k)}(t+1)=i}\\
        E^{(k)}(t+1) &=
        \begin{cases}
        \mathrm{min}(\omega_{\mathrm{max}}, E^{(k)}(t) + \Delta),\\
        \qquad \qquad \qquad \qquad \mbox{if }\mathrm{owner}(k,t)\\
        \mathrm{max}(\omega_{\mathrm{min}},E^{(k)}(t) - \Delta),\\
        \qquad \qquad \qquad \qquad\mbox{if }\invneg \mathrm{owner}(k,t)
        \end{cases}\\
        S^{(k)}(t+1) &= \indicator{E^{(k)}(t+1) = \omega_{\mathrm{min}}}
      \end{array} \right.
  \end{split}
\end{align}

The first equation of system $\phi$ is responsible for moving each particle to a new node $j$, where $j$ is determined according
to the time-varying transition matrix in (\ref{eq:transition-matrix-modified}). In other words, the acquisition of $p(t+1)$ is performed by generating random numbers following the distribution of the transition matrix $\mathbb{P}^{(k)}_{\mathrm{transition}}(t)$. The second equation updates the number of visits that node $i$ has received by particle $k$ up to time $t$;
the third equation is used to maintain the current energy levels of all the particles inserted in the network; and the fourth
equation indicates whether the particle is active or exhausted, depending on its actual energy level. Note that system $\phi$ is nonlinear. This occurs on account of the indicator function, which is nonlinear. One can also see that system $\phi$ is Markovian, since the future state only depends on the present state.

\section{Results and Discussion}
\label{resultados}

\Red{In this section, we present synthetic examples with the goal of elucidating how the particle competition technique works. Next, we apply it to a real-world application of authors' names disambiguation. With regard to the technique's parameter selection, the  guidelines proposed in \cite{Thiago2012a} are followed. Hence, we will use $\lambda = 0.6$, $\epsilon = 0.05$, and $[\omega_{\mathrm{min}}, \omega_{\mathrm{max}}] = [0, 1]$. The calibration of $K$ depends on the type of the data set which we are dealing with. For its estimation, we also utilize the heuristic presented in \cite{Thiago2012a}.}

\subsection{Simulation on a Synthetic Data Set}
\label{Simulation-Numerical-Examples}

\begin{figure}
    \centering
    \includegraphics[scale = 0.24]{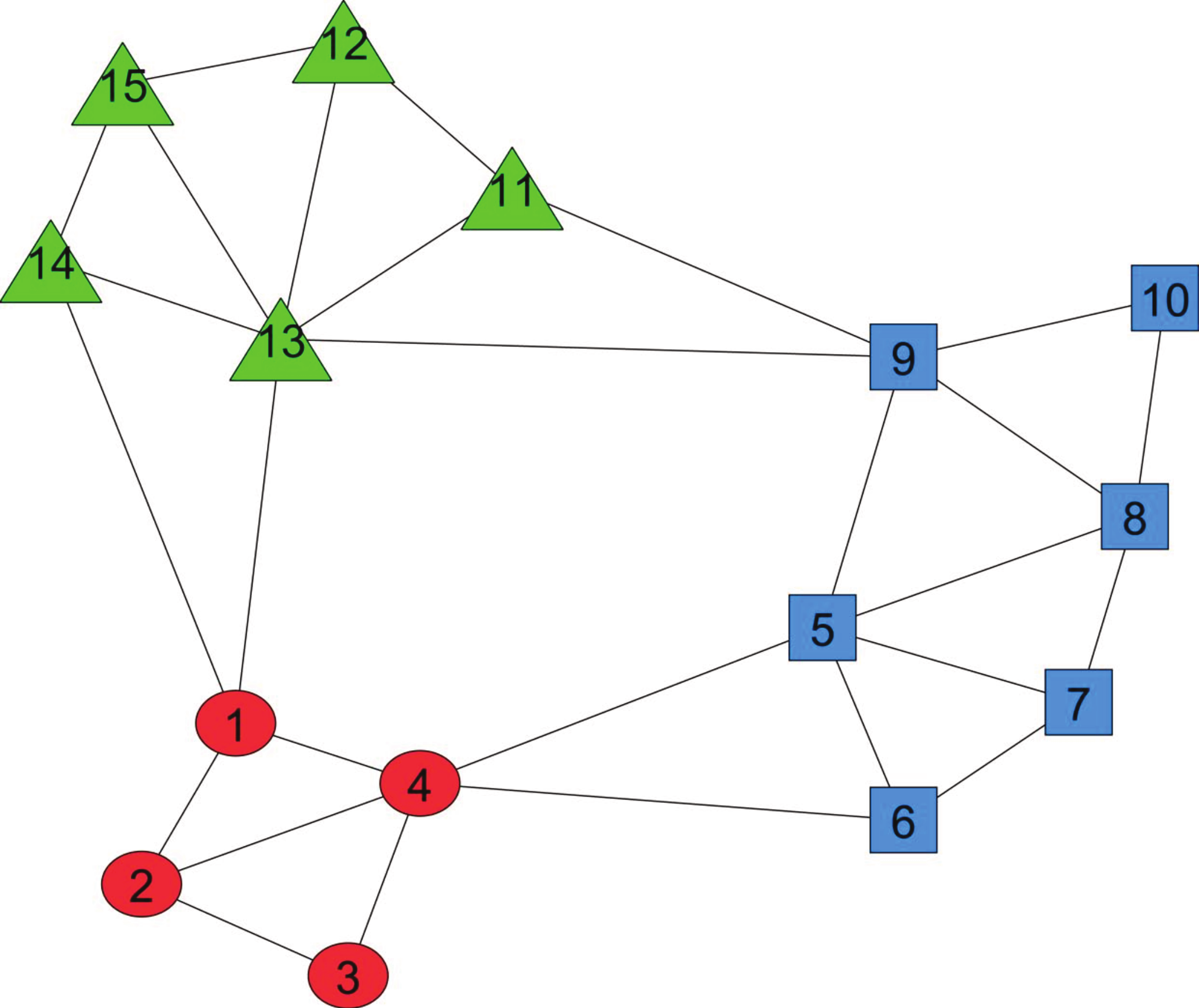}
    \caption{A simple networked data set. The red or ``circle'' group is composed by the nodes $1$ to $4$, the blue or ``square'' group comprises the nodes numbered from $5$ to $10$, and the green or ``triangle'' class encompasses the nodes $11$ to $15$. The colorful nodes have only been drawn for illustrative purposes. In the unsupervised task, all the external information is ignored.}
    \label{fig:simple-example}
\end{figure}

\begin{figure*}
    \centering
    \subfloat[]
    {\includegraphics[scale = 0.19]{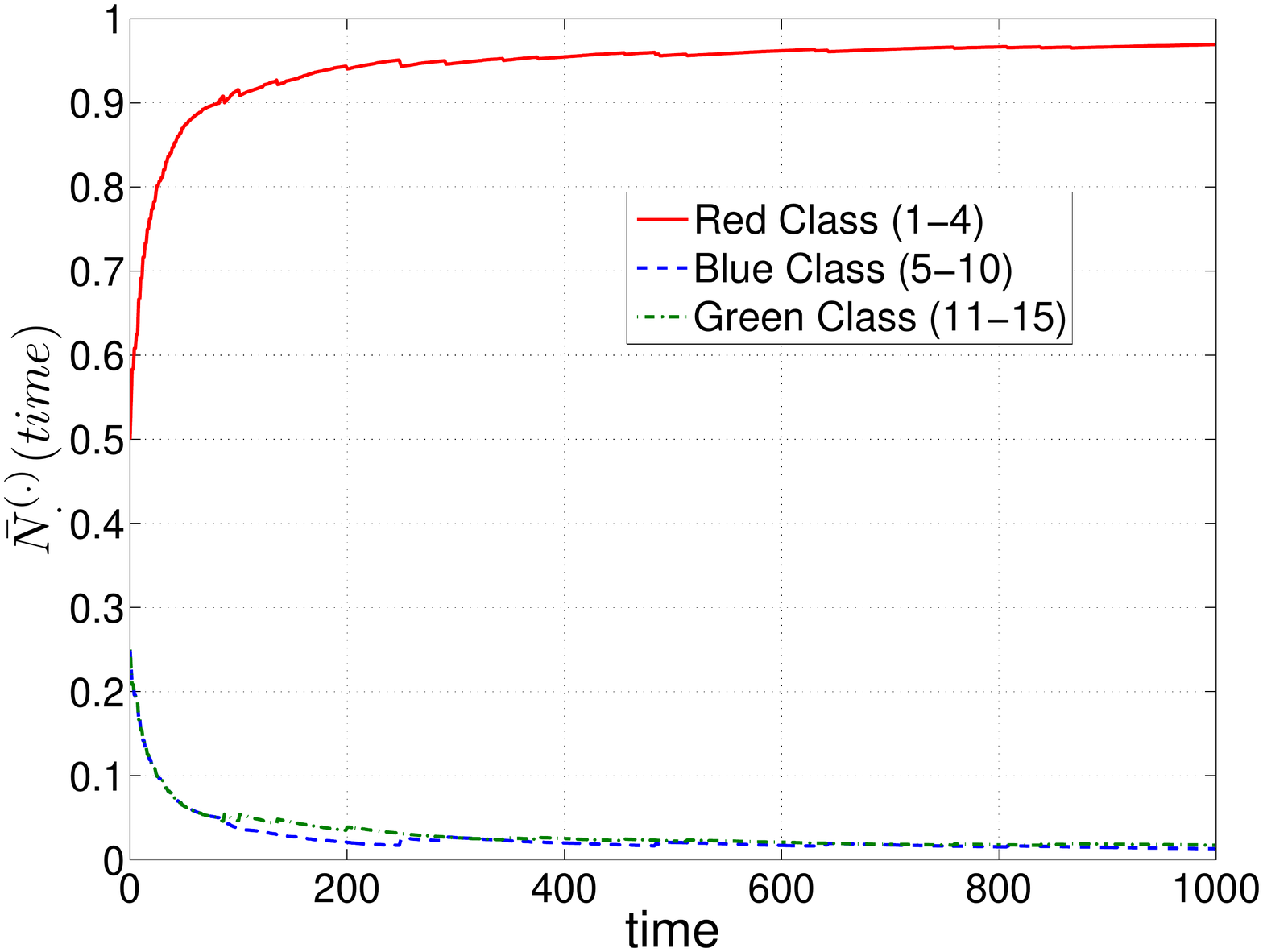}\label{fig:dominationCom1}}
    \subfloat[]
    {\includegraphics[scale = 0.19]{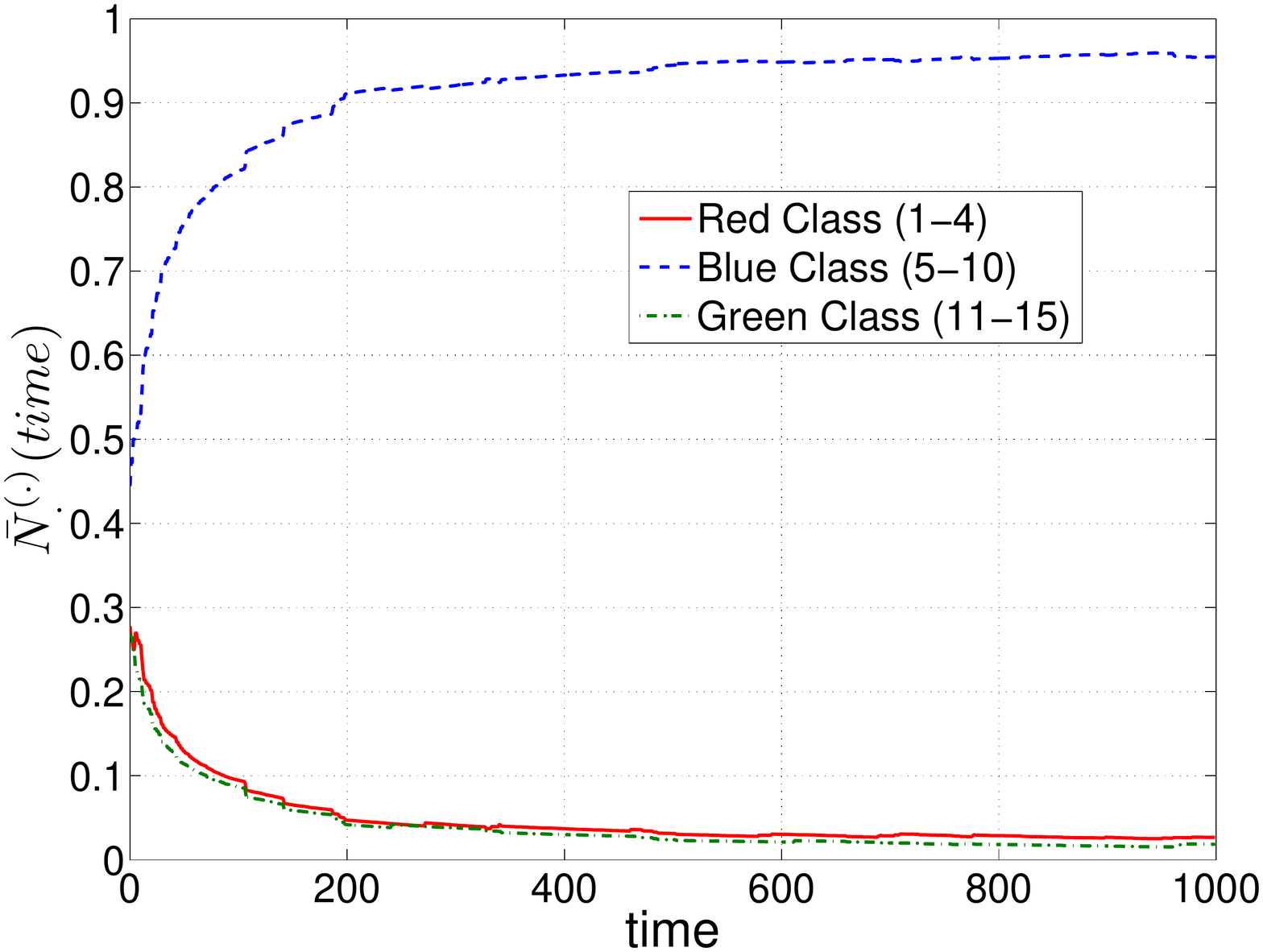}\label{fig:dominationCom2}}
    \subfloat[]
    {\includegraphics[scale = 0.19]{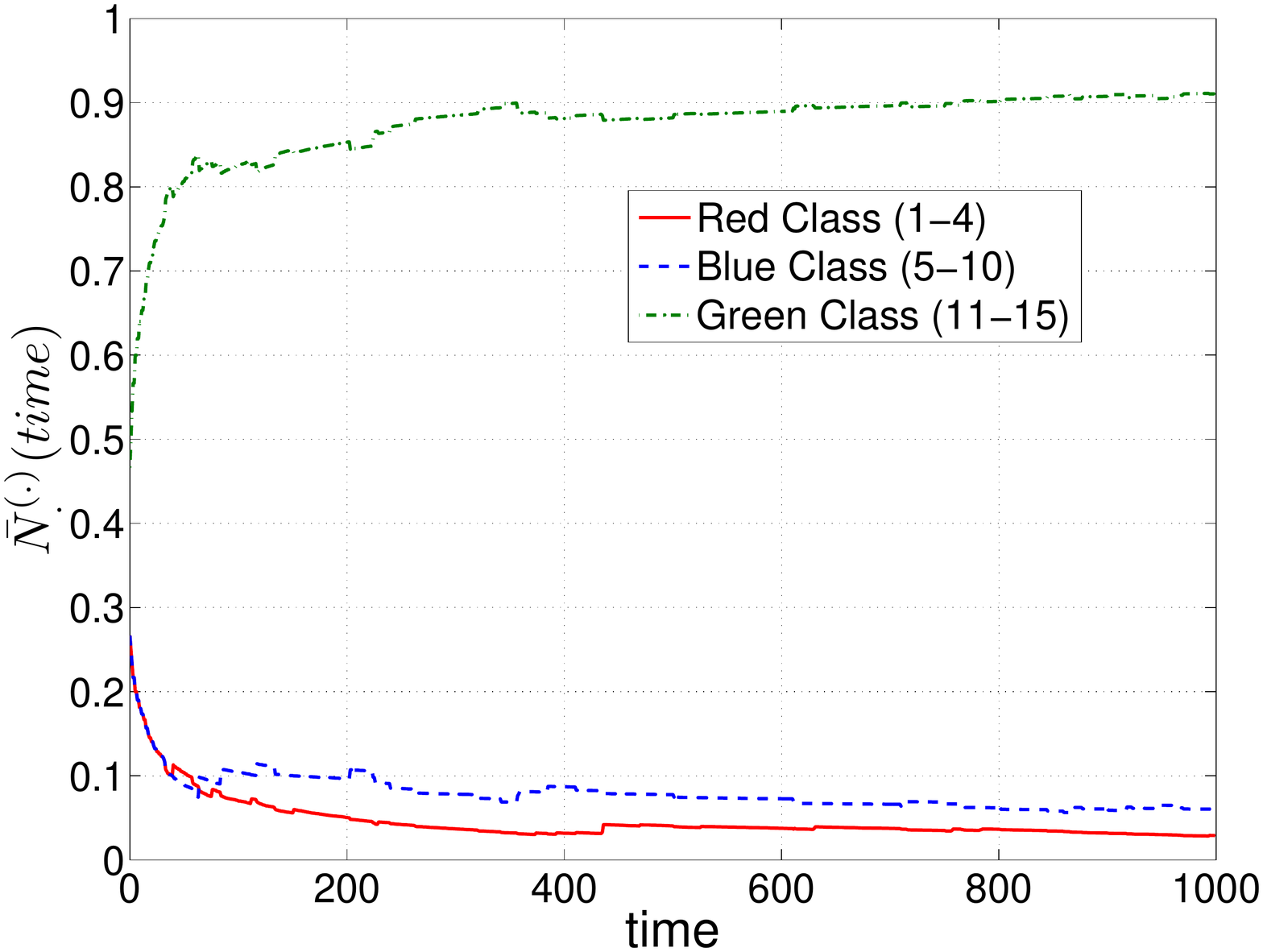}\label{fig:dominationCom3}}
    \caption{Evolutional behavior of the average class domination level imposed by the 3 particles in the network on: (a) the red or ``circle'' class, (b) the blue or ``square'' class, and (c) the green or ``triangle'' class.}
\end{figure*}

\Red{In this section, we provide a simple computer simulation on a networked synthetic data set with the purpose of illustrating how the proposed algorithm works. Specifically, the temporal evolution of matrix $\bar{N}(t)$ for a network consisting of $V=15$ nodes split into $3$ unbalanced communities, as depicted in Fig. \ref{fig:simple-example}, is analyzed. $K=3$ particles are inserted into the network at the initial positions $p(0) = [2 \ 4\ 13]$, meaning the first particle starts at node $2$, the second particle starts at node $4$, and the third particle starts at node $13$. The competitive system is iterated until $t=1000$ and the predicted label for each of the unlabeled nodes is given by the particle's label that is imposing the highest domination level. Figures \ref{fig:dominationCom1}, \ref{fig:dominationCom2}, and \ref{fig:dominationCom3} show the evolutional behavior of the domination levels imposed by the three particles on the red or ``circle'' community, the blue or ``square'' community, and the green or ``triangle'' community, respectively. Specifically, from Fig. \ref{fig:dominationCom1}, we can verify that red or ``circle'' particle dominates
nodes $1$ to $4$ (red or ``circle'' community), due to the fact that the average domination level on these nodes approaches $1$, whereas the average domination
levels of the other two rival particles decay to $0$. Considering Figs. \ref{fig:dominationCom2} and \ref{fig:dominationCom3}, we can use the same logic to confirm that the blue or ``square'' particle completely dominates the nodes $5$ to $10$ (blue or ``square'' community) and the green or ``triangle'' particle dominates nodes $11$ to $15$ (green or ``triangle'' community). In order to check the particles' initial locations independence, we have purposefully put all the particles starting from the node $2$. Again, we have verified that the particle competition model has discovered all the communities in a correct manner.}

\section{Disambiguating authors' names} \label{resultado2}

To assess the efficiency of the algorithm based on the unsupervised competitive learning we apply the algorithm to a set of ambiguous authors publishing  preprints on the arXiv repository. In special, we used the same database reported in previous investigations~\cite{amancio1,amancio2}. The data are divided according to the number $\eta$ of different authors with ambiguous name. For each value of $\eta = \{2,3,4,5,6,7,8,9\}$ we computed the quality of the disambiguation using the so called $f$-measure~\cite{fmeasure,SI} based on both precision~\cite{fmeasure} and recall~\cite{fmeasure} of the partitions. The results obtained for $\eta = 2$, $\eta = 3$,  $\eta = 5$ and $\eta = 9$ are shown in the first column of Tables \ref{t1}, \ref{t2}, \ref{t3} and \ref{t4}, respectively~\cite{SI}. These results were compared with traditional algorithms, where entities with ambiguous names are represented by a vector $\overrightarrow{v}$ so that each element $i$ of $\overrightarrow{v}$ represents the presence or absence of the author $i$ as a neighbor. In other words, if $i$ appeared as a neighbor of the homonymous author, then $\overrightarrow{v}(i)$ = 1. Otherwise, $\overrightarrow{v}(i)$ = 0. In this case, the clustering was performed with a set of competing techniques.\Red{
Note that all parameters were employed according to their original papers~\cite{Dempster1977,KMeans1977,Khan2004,Ward1963,Karypis2003,Clauset2004}. The techniques are given as follows:}

\begin{itemize}
    \item \Red{Partitional algorithms: Expectation Maximization~\cite{Dempster1977} (EM) and K-Means with optimized center initialization~\cite{Khan2004};}

    \item \Red{Hierarchical algorithms: CHAMELEON \cite{Karypis2003} (an agglomerative graph-based technique), Modularity Greedy algorithm \cite{Clauset2004} (also an agglomerative graph-based method) and Wards~\cite{Ward1963}.}
\end{itemize}

\Red{The best result among these five algorithms is also shown in the second column of Tables \ref{t1}, \ref{t2}, \ref{t3} and \ref{t4}. Note that, consistently, the $f$-measure tends to decrease in all algorithms as more ambiguous names are introduced in the network. Nevertheless, the approach based on competitive learning outperforms the competing techniques in most of the databases. In order to check the significance of these results, we calculate the value $p$-value representing the probability that the competitive learning technique outperforms the competing algorithms just by chance $N$ or more times as}~\cite{ross}
\begin{equation}
    p(N) = \sum_{n=N}^{10} \binom{10}{n} \Bigg{[} \frac{1}{6} \Bigg{]}^{n} \Bigg{[} \frac{5}{6} \Bigg{]}^{10 - n}.
\end{equation}
\Red{Table \ref{t5} confirms the significance of the results because all $p$-values are lower than $1.5 \times 10^{-2}$}.

\begin{table}
\caption{\label{t1} $f$-measure obtained with the algorithm based on particles (first column) and with traditional algorithms based on the recurrence of the neighbors (second column). All 10 databases contain 2 authors with the same name.
The best $f$-measure achieved in each data set is bolded. In most cases the approach based on particles outperforms the traditional approach.}
\begin{ruledtabular}
\begin{tabular}{cccc}
DB & Particles & Best Traditional & Algorithm \\
\hline
A & ${\bf 0.984 \pm 0.028}$ & 0.973 & EM \\
B & ${\bf 0.974 \pm 0.021}$ & 0.969 & CHAMELEON \\
C & $0.953 \pm 0.040$ & {\bf 0.976} & EM \\
D & $0.953 \pm 0.048$ & {\bf 0.974} & CHAMELEON\\
E & ${\bf 0.893 \pm 0.011}$ & 0.979 & CHAMELEON \\
F & ${\bf 0.900 \pm 0.028}$ & 0.873 & EM\\
G & ${\bf 0.820 \pm 0.012}$ & 0.807 & Modularity\\
H & ${\bf 0.824 \pm 0.026}$ & 0.699 & Ward\\
I & ${\bf 0.824 \pm 0.004}$ & 0.681 & K-Means\\
J & $0.754 \pm 0.034$ & {\bf 0.778} & Modularity\\
\end{tabular}
\end{ruledtabular}
\end{table}

\begin{table}
\caption{\label{t2} $f$-measure obtained with the algorithm based on particles (first column) and with traditional algorithms based on the recurrence of the neighbors (second column). All 10 databases contain 3 authors with the same name.
The best $f$-measure achieved in each data set is bolded. In most cases the approach based on particles outperforms the traditional approach.}
\begin{ruledtabular}
\begin{tabular}{cccc}
DB & Particles & Best Traditional & Algorithm \\
\hline
A & ${\bf 0.838 \pm 0.051}$ & 0.829 & CHAMELEON  \\
B & $0.819 \pm 0.020$ & {\bf 0.826} & CHAMELEON  \\
C & ${\bf 0.789 \pm 0.015}$ & 0.717 & Modularity \\
D & ${\bf 0.759 \pm 0.052}$ & 0.741 & CHAMELEON  \\
E & ${\bf 0.739 \pm 0.031}$ & 0.723 & EM \\
F & ${\bf 0.729 \pm 0.028}$ & 0.711 & EM \\
G & ${\bf 0.719 \pm 0.030}$ & 0.705 & Modularity \\
H & ${\bf 0.710 \pm 0.029}$ & 0.692 & EM \\
I & ${\bf 0.689 \pm 0.029}$ & 0.640 & Ward \\
J & ${\bf 0.670 \pm 0.025}$ & 0.612 & Ward \\
\end{tabular}
\end{ruledtabular}
\end{table}

\begin{table}
\caption{\label{t3} $f$-measure obtained with the algorithm based on particles (first column) and with traditional algorithms based on the recurrence of the neighbors (second column). All 10 databases contain 5 authors with the same name. The best $f$-measure achieved in each data set is bolded. In most cases the approach based on particles outperforms the traditional approach.}
\begin{ruledtabular}
\begin{tabular}{cccc}
DB & Particles & Best Traditional & Algorithm \\
\hline
A & ${\bf 0.859 \pm 0.051}$ & 0.716 & EM \\
B & ${\bf 0.719 \pm 0.068}$ & 0.620 & Ward \\
C & ${\bf 0.669 \pm 0.113}$ & 0.642 & Modularity \\
D & ${\bf 0.660 \pm 0.024}$ & 0.576 & EM \\
E & ${\bf 0.649 \pm 0.049}$ & 0.614 & K-Means  \\
F & $0.620 \pm 0.034$ & {\bf 0.709} & CHAMELEON  \\
G & ${\bf 0.590 \pm 0.066}$ & 0.589 & CHAMELEON  \\
H & ${\bf 0.561 \pm 0.028}$ & 0.548 & CHAMELEON \\
I & ${\bf 0.508 \pm 0.085}$ & 0.490 & Modularity \\
J & $0.489 \pm 0.041$ & {\bf 0.545} & Ward \\
\end{tabular}
\end{ruledtabular}
\end{table}

\begin{table}
\caption{\label{t4} $f$-measure obtained with the algorithm based on particles (first column) and with traditional algorithms based on the recurrence of the neighbors (second column). All 10 databases contain 9 authors with the same name. The best $f$-measure achieved in each data set is bolded. In most cases the approach based on particles outperforms the traditional approach.}
\begin{ruledtabular}
\begin{tabular}{cccc}
DB & Particles & Best Traditional & Algorithm \\
\hline
A & ${\bf 0.739 \pm 0.019}$ & 0.599 & Ward \\
B & ${\bf 0.638 \pm 0.031}$ & 0.593 & CHAMELEON \\
C & ${\bf 0.608 \pm 0.039}$ & 0.574 & CHAMELEON \\
D & ${\bf 0.590 \pm 0.010}$ & 0.519 & CHAMELEON \\
E & $0.588 \pm 0.019$ & {\bf 0.589} & Modularity  \\
F & ${\bf 0.581 \pm 0.039}$ & 0.577 & CHAMELEON  \\
G & $0.532 \pm 0.024$ & {\bf 0.537} & Ward  \\
H & $0.527 \pm 0.036$ & {\bf 0.553} & Modularity \\
I & $0.479 \pm 0.069$ & {\bf 0.524} & CHAMELEON \\
J & $0.458 \pm 0.023$ & {\bf 0.528} & CHAMELEON \\
\end{tabular}
\end{ruledtabular}
\end{table}

\begin{table}
\caption{\label{t5} $p$-value representing the likelihood of the proposed algorithm to perform better than the other three traditional algorithms just by chance. Note that in all cases the values are significative, which confirms the efficiency of the algorithm.}
\begin{ruledtabular}
\begin{tabular}{cccc}
Number of ambiguous authors & $p$-value \\
\hline
2 ambiguous authors &  \Red{$2.7 \times 10^{-4}$} \\
3 ambiguous authors &  \Red{$1.9 \times 10^{-5}$} \\
4 ambiguous authors &  \Red{$2.7 \times 10^{-4}$} \\
5 ambiguous authors &  \Red{$1.9 \times 10^{-5}$} \\
6 ambiguous authors &  \Red{$2.7 \times 10^{-4}$} \\
7 ambiguous authors &  \Red{$1.5 \times 10^{-2}$} \\
8 ambiguous authors &  \Red{$2.7 \times 10^{-4}$} \\
9 ambiguous authors &  \Red{$1.5 \times 10^{-2}$} \\
\end{tabular}
\end{ruledtabular}
\end{table}

One can wonder the reason behind the proposed technique is more suitable to disambiguate names in collaborative networks than traditional algorithms. The competitive process performed by the particles in the network is able to capture the topological features of the data by using the links of the network. \Red{Since our network formation step is composed of a linear combination of random walks with varying lengths, where we strengthen the relationships of similar data and weaken the relationship of different data by simply adjusting the edge weight of each pair of nodes, we expect the resulting network to reliably reflect the characteristics of the collaborative network. Now, using this representative network, a set of particles is put into the nodes of the network. These particles navigate into the network with the purpose of dominating new vertices by constantly visiting them. Simultaneously, the particles attempt to reject intruder particles indirectly through their energy levels and also through the reanimation procedure embedded within the method. That is, whenever particles are visiting vertices dominated by rival particles, they suffer a loss in their energy levels. Eventually, they become exhausted if they continuously visit these kinds of vertices.
Therefore, this mechanism serves as a repulsive force to maintain stability among the territories (subset of dominated vertices) of different particles. Additionally, the particles move in the network according to two orthogonal dependencies: defensive and exploratory approaches. Since both approaches are nonlinear, we expect that the particles will be able to discover communities of both regular or irregular forms.}
Given that the network can represent arbitrary forms of data distributions, our network-based model is able to provide better community detection accuracies.  In contrast to that, traditional techniques often rely on assumptions pervading the distribution of the data items, which, in turn, may be infeasible to estimate in some situations, such as in the problem here tackled. Hence, they may not perform well in these situations.

\section{Conclusion}
\label{conclusao}

The term ambiguity refers to the ability of expression conveying at least two possible interpretations in the absence of contextual information. This phenomenon occurs in many situations of scientific interest and particularly in the representation of authorship in scientific papers. In the current study, we treated the problem of disambiguating authors' names by introducing a novel network-based methodology. \Red{We motivate the use of a networked environment over vector space data because of the fact that networks are able to capture the topology of the data relationships and, hence, is able to enhance the learning process of machine learning techniques. Furthermore, there is no weight between authors in the vector space approach, while there is weight in the graph representation. This permits a natural and intuitive way of representing more similar connections between different authors than in a vector-based approach.}

In the proposed method, after building a collaborative network, we applied a technique based on the dynamics of particles walking on the collaborative network according to rules determined by an hybrid walk based on random and preferential factors. Interestingly, the proposed methodology turned out to be useful to discriminate authors' names in the unsupervised scheme, as a significant improvement of the task was observed when we compared our technique with the traditional methods. Because the strategy is generic, we intend to study its applicability to a series of other problems related to the disambiguation of generic entities. More specifically, we intend to extend it to natural language processing tasks such as in the problem of word sense disambiguation.

\begin{acknowledgments}
    Thiago C. Silva (2009/12329-1) and Diego R. Amancio (2010/00927-9) acknowledge the financial support from FAPESP.
\end{acknowledgments}


\providecommand{\noopsort}[1]{}\providecommand{\singleletter}[1]{#1}%

\end{document}